\documentclass[aps,pra,twocolumn,10pt,bibnotes]{revtex4-1} 

\usepackage[pdftex]{graphicx}
\usepackage{graphicx}
\usepackage{amsmath,ulem}
\usepackage{color}
\usepackage{lineno}
\usepackage{epstopdf}

\usepackage{textcomp}
\usepackage{units}
\usepackage{verbatim}
\usepackage{scrextend}

\def\41K{$^{41}$K} 
\def\39K{$^{39}$K}
\def\87Rb{$^{87}$Rb}

\def\EE#1{\times 10^{#1}}
\def\ket#1{\left|#1\right\rangle}
\newcommand{\folder}{}

\begin{document}
\title{Tunable dual-species Bose-Einstein condensates of \39K and \87Rb}

\author{L. Wacker$^1$, N. B. J\o rgensen$^1$, D. Birkmose$^1$, R. Horchani$^1$, W.~Ertmer$^2$, C.~Klempt$^2$, N.~Winter$^1$, J.~Sherson$^1$, J.J.~Arlt$^1$}

\affiliation{$^1$ Institut for Fysik og Astronomi, Aarhus Universitet, Ny Munkegade 120, DK-8000 Aarhus C}
\affiliation{$^2$ Institut f\"ur Quantenoptik, Leibniz Universit\"at Hannover, Welfengarten~1, D-30167~Hannover}

\date{\today}

\begin{abstract}
We present the production of dual-species Bose-Einstein condensates of \39K and \87Rb. Preparation of both species in the $\ket{F=1,m_F=-1}$~state enabled us to exploit a total of three Fesh\-bach resonances which allows for simultaneous Feshbach tuning of the \39K intraspecies and the \39K-\87Rb interspecies scattering length. Thus dual-species Bose-Einstein condensates were produced by sympathetic cooling of \39K with \87Rb. A dark spontaneous force optical trap was used for \87Rb, to reduce the losses in \39K due to light-assisted collisions in the optical trapping phase, which can be of benefit for other dual-species experiments. The tunability of the scattering length was used to perform precision spectroscopy of the interspecies Feshbach resonance located at \unit[117.56(2)]{G} and to determine the width of the resonance to \unit[1.21(5)]{G} by rethermalization measurements. The transition region from miscible to immiscible dual-species condensates was investigated and the interspecies background scattering length was determined to $\unit[28.5]{a_0}$ using an empirical model. This paves the way for dual-species experiments with \39K and \87Rb BECs ranging from molecular physics to precision metrology.
\end{abstract}

\maketitle

\section{Introduction}

Within the past decade, research on ultracold atoms has moved from the investigation of their fundamental properties to the application of ultracold samples in quantum simulation and precision metrology. The ability to tailor external potentials freely and to manipulate the interaction strength within the samples has led to numerous advances in the field. In particular mixed quantum gases have attracted considerable interest, since they offer a wealth of research opportunities. These include the creation of deeply bound dipolar molecules~\cite{Carr2009}, the investigation of few-particle physics~\cite{Zinner2014,Zinner2013}, the observation of quantum phases in optical lattices~\cite{Bloch2005} and precision measurements~\cite{Schlippert2014}.

Such mixed quantum gasses can generally be realized by using a single atomic species in multiple quantum states, by using multiple isotopes of the same species, or by using different atomic species. Thus it is possible to realize Bose-Fermi, Bose-Bose or Fermi-Fermi mixtures. Since cooling techniques to achieve ultracold temperatures have become available for an increasing number of atomic species, this leads to a considerable number of possible mixtures. Depending on the experimental requirements different mixtures may be optimal, e.g. only some mixtures lend themselves to the production of dipolar ground state molecules due to their chemical stability. 

In particular, dual-species Bose-Einstein condensates have to date been studied by investigating $^{87}$Rb-$^{41}$K~\cite{Modugno2002,Thalhammer2008}, $^{87}$Rb-$^{133}$Cs~\cite{McCarron2011,Lercher2011}, $^{23}$Na-$^{87}$Rb~\cite{Xiong2013}, $^{87}$Rb-$^{84}$Sr, $^{87}$Rb-$^{88}$Sr~\cite{Pasquiou2013}, and isotopic mixtures were investigated using $^{87}$Rb-$^{85}$Rb~\cite{Papp2008}, $^{168}$Yb-$^{174}$Yb~\cite{Sugawa2011}. Key achievements with these mixtures have included control of the interspecies interactions~\cite{Thalhammer2008}, the observation of heteronuclear Efimov resonances~\cite{Barontini2009}, phase separation between the components~\cite{McCarron2011}, the creation of deeply bound molecules~\cite{Takekoshi2014,Molony2014} and the observation of heteronuclear spin dynamics~\cite{Li2015}. 

\begin{figure}[tbp]
	\centering
	\includegraphics[width=8.6cm]{\folder 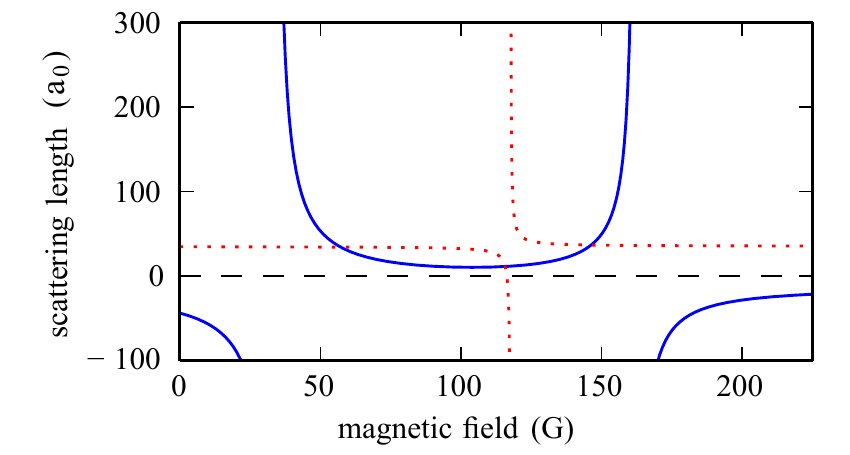}
	\caption{Intra- and interspecies scattering lengths for \39K and \87Rb in the $\ket{F=1,m_f=-1}$~state. The intraspecies scattering length of \39K~(solid blue line) is positive between two Feshbach resonances at \unit[33.6]{G} and \unit[162.35]{G}. Within this window the interpecies scattering length can be tuned by a Feshbach resonance at \unit[117.56]{G} (red dotted line).}
	\label{fig:Scattering}
\end{figure}

Here we describe the production of a dual-species Bose-Einstein condensate (BEC) of \39K and \87Rb with tunable interactions. This mixture is experimentally well accessible, since \39K does not require any specialized atom sources, as it is the most abundant potassium isotope and the necessary Feshbach resonances are located at moderate fields. However, \39K is typically not considered as a useful constituent in dual-species experiments, since it has a negative background scattering length. Thus magnetic field tuning of the scattering length is necessary to achieve Bose-Einstein condensation of \39K and hence the magnetic field can typically not be chosen freely to tune the intraspecies scattering length. 

This general statement is however not true when both \39K and \87Rb are in the $\ket{F=1,m_F=-1}$~state as shown in Fig.~\ref{fig:Scattering}. In this state combination, two intraspecies Fesh\-bach resonances~\cite{DErrico2007} produce a $\approx\unit[130]{G}$ wide magnetic field window where the \39K scattering length is positive. Within this window an interspecies Feshbach resonance at \unit[117.56]{G}~\cite{Simoni2008} allows the interactions between \39K and \87Rb to be tuned freely. We produce dual-species Bose-Einstein condensates on both sides of this Feshbach resonance by sympathetic cooling of \39K with \87Rb in a crossed dipole trap.  To demonstrate the tunablility of the interactions we perform Feshbach spectroscopy with ultracold samples and investigate the transition from miscible to immiscible dual-species BECs. 

This paper is structured as follows. Section~\ref{Setup} describes the production of the dual-species BEC. This includes the dual atom collection in section~\ref{MOT} with particular emphasis on the performance of a dark spontaneous optical force trap for \87Rb. Section~\ref{magnetictrap} describes the magnetic trapping and initial sympathetic cooling of both species while section~\ref{dipoletrap} details the optical trapping and state preparation procedure. Feshbach tuning and the creation of dual-species BECs is characterized in section~\ref{dualBEC}. Finally our results on Feshbach spectroscopy and phase-separation are provided in section~\ref{interactions}.

\section{Dual-species Bose-Einstein condensation}
\label{Setup}

The experimental apparatus employed here was previously used to realize an atom laser~\cite{KleineBuning2010} and to perform precision measurements~\cite{KleineBuning2011} based on \87Rb alone. Briefly, the collection region (see Fig.~\ref{fig:dark_spot}) consists of a large glass cell to enable efficient loading of a magneto-optical trap (MOT) for \39K and a dark spontaneous optical force trap (dark-SPOT)~\cite{Ketterle1993} for \87Rb. The \39K and \87Rb atoms are provided by commercial dispenser sources. In addition, light-induced atom desorption~\cite{Klempt2006} is used to desorb atoms from the surface of the glass cell.

Subsequent experiments on dual-condensates are performed in a small L-shaped glass cell (science cell). The two cells are separated by a differential pumping stage which allows for long lifetimes of the atomic samples in the science cell. The transport of the atoms between the cells is realized with a movable magnetic quadrupole trap. The science cell is designed in an L-shape to provide good optical access from all six spatial directions.  

\subsection{Simultaneous magneto-optical trapping of \39K and \87Rb} 
\label{MOT}

The simultaneous collection of \39K and \87Rb in spatially overlapping magneto-optical traps poses two interrelated problems. First of all, \39K is not particularly well suited for laser cooling due to its small excited state hyperfine splitting~\cite{Kishimoto2009, Landini2012,Gokhroo2011}. Thus, only relatively small samples can be collected and special strategies have to be employed to reach low temperatures~\cite{Landini2012}. This problem is aggravated by the interaction of the large \87Rb samples with small \39K samples. In particular, light-assisted collisions lead to similar atom losses from both samples~\cite{Marcassa2000}, which have a larger relative effect on the small \39K samples.

\begin{figure}[htbp]
	\centering
		\includegraphics[width=8cm]{\folder 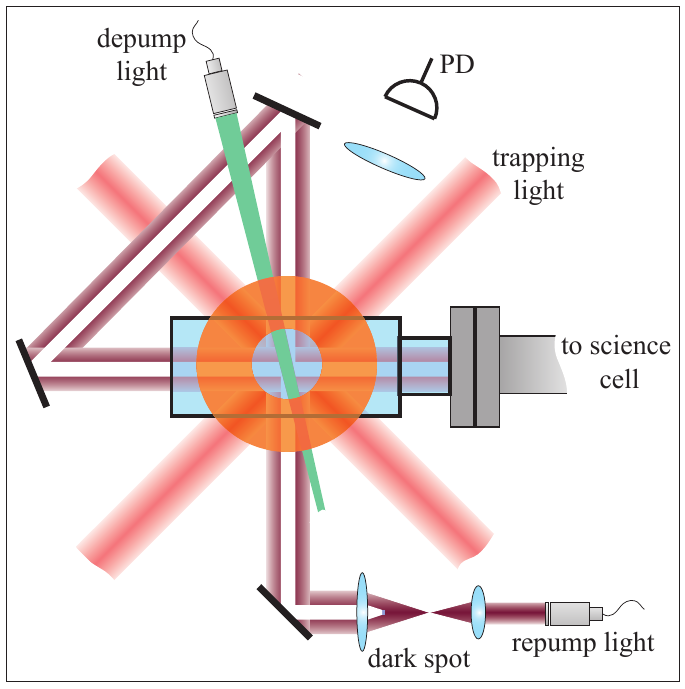}
	\caption{Sketch of the collection region (horizontal plane) including the \39K MOT and the \87Rb dark-SPOT. The MOT beams (red) contain trapping light for both species and repumping light for \39K. The repumping (dark red) and depumping light (green) for \87Rb is applied separately. The repuming light contains a dark spot of adjustable size (see text). The glass cell (blue) and the quadrupole coils (orange) are indicated.}
	\label{fig:dark_spot}
\end{figure}

We overcome these problems by using a \39K MOT with large trapping laser beams in combination with a dark-SPOT~\cite{Ketterle1993} for \87Rb as shown in Fig.~\ref{fig:dark_spot}. Due to the dark-SPOT, \87Rb atoms at the center of the dual trap accumulate in the $\ket{F=1}$ state and hence light-assisted collisions are avoided. Note that this represents a new application of the dark-SPOT technique, which was previously used primarily to increase the phase-space density in single species experiments~\cite{Ketterle1993,Anderson1994,Townsend1996,Radwell2013}.
	
To operate the \39K MOT, a single optical fiber delivers \unit[200]{mW} of light detuned \unit[24]{MHz} below the transition from the ground state~$\ket{F=2}$ to the excited state~$\ket{F=3'}$ and \unit[200]{mW} detuned \unit[32]{MHz} below the $\ket{1}$ to $\ket{2'}$ transition. In the following, these are called trapping and repumping light respectively, despite the fact that their roles are not clearly distinct. This light is split into six beams using achromatic beam splitters. Galilean telescopes are used to magnify the beams to a $1/\mathrm{e}^2$ diameter of \unit[34]{mm} before intersecting in the MOT region. These large beams allow for the efficient accumulation of atoms from the background vapor. In addition, \unit[280]{mW} of trapping light for \87Rb tuned \unit[24]{MHz} below the $\ket{2}$ to $\ket{3'}$~transition is delivered by this fiber.

As shown in Fig.~\ref{fig:dark_spot} repumping light on the $\ket{1}$ to $\ket{2'}$ transition and depumping light on the $\ket{2}$ to $\ket{2'}$ transition are delivered in separate fibers to realize the dark-SPOT for \87Rb. The repumping light is first collimated and then magnified to a beam diameter of \unit[22]{mm} in a 2:1 telescope. The dark spot in the beam is realized by placing a small opaque disk of \unit[6]{mm} diameter within this telescope. Thus the effective spot size can be varied by translating the disk. Due to its large size this dark spot propagates without considerable diffraction over the distance required in the experiment. To obtain a central dark region of the trap surrounded by regions with repumping light, the beam is recycled and passes the trap for a second time in an orthogonal direction (see Fig.~\ref{fig:dark_spot}). Additional depumping light is required in this region, since most atoms would otherwise remain in the cooling cycle while traversing it. Note, that this simple setup can easily be implemented in numerous multi species experiments to avoid losses due to light-assisted collisions.

The fluorescence light is collected on a photo diode to quantify the number of atoms acquired in the MOT. We distinguish the weak \39K signal from the large \87Rb signal by turning off all light for the \87Rb MOT and tuning the \39K light to resonance.

\begin{figure}[htbp]
		\centering
	  \includegraphics[width=8.6cm]{\folder 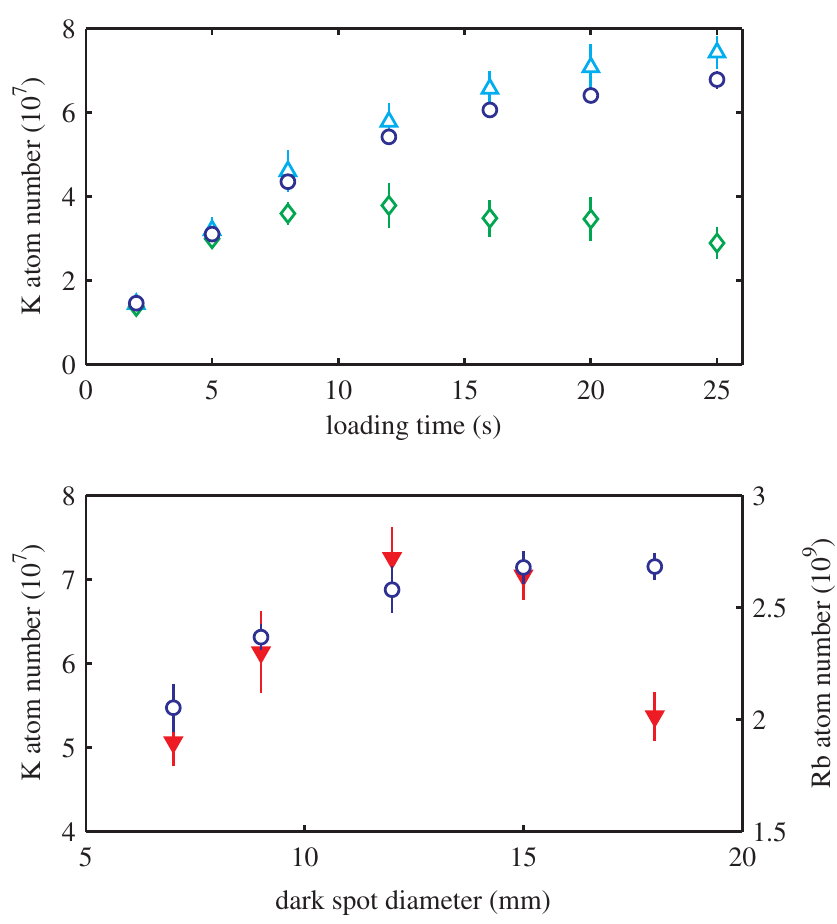}
	\caption{Performance of the combined \39K MOT and \87Rb dark-SPOT. (top) Atom number in a single species \39K MOT (blue triangles), a \39K MOT in the presence of a bright \87Rb MOT (green diamonds) and a \39K MOT in the presence of a \87Rb dark-SPOT (dark blue circles). Simultaneous loading of a bright \87Rb MOT leads to a clear loss of \39K atoms, while almost no loss is visible in the dark-SPOT case. (bottom) Atom number in the \39K MOT (dark blue circles) and the \87Rb dark-SPOT (red triangles) as a function of the dark spot diameter.}
	\label{fig:MOT_loading}
\end{figure}

Two experiments are performed to characterize the performance of the dual-species optical trap. Figure~\ref{fig:MOT_loading} shows the \39K atom number during MOT loading under different experimental conditions. The maximal number of \39K atoms is obtained by loading a single species MOT. In the presence of a bright \87Rb MOT (realized by removing the opaque disk) severe losses lead to an inferior steady state with less than half of the number of atoms compared to the single-species case. By using the dark-SPOT for \87Rb these losses are avoided and one obtains almost the same number of \39K atoms as in the single species MOT.

The size of the dark spot was optimized with respect to both species by measuring the atom numbers after \unit[25]{s} loading time as a function of the spot size (see Fig.~\ref{fig:MOT_loading}). Thus an effective spot diameter of \unit[12]{mm} was chosen. For smaller sizes, the \87Rb atoms are not depumped properly in the area of the \39K MOT, resulting in losses. Larger sizes lead to a lower number of \87Rb atoms, while no further gain of \39K atoms is obtained. Contrary to previous experiments~\cite{Radwell2013}, no decrease in the \87Rb loading rate was observed for this dark spot size. Thus, up to $7\EE7$ \39K and $2.7\EE9$ \87Rb atoms can be collected. In practice, the loading time of \39K is adjusted to obtain an appropriate relative atom number for subsequent sympathetic cooling. 

After loading sufficient atoms in the dual-species optical trap, an optical molasses is applied to cool the atoms well below the Doppler limit. A typical optical molasses works efficiently, if the excited state hyperfine splitting is much larger than the natural linewidth $\Gamma$, which is the case for rubidium. In our experiments, a molasses for \87Rb atoms is realized by turning off the magnetic field gradient and detuning the trapping light by $-7~\Gamma$ for a duration of \unit[9]{ms}. For \39K, the situation is considerably different. Since the hyperfine splitting is on the same order as the linewidth, a standard molasses procedure is not efficient. However, it has been shown that sub-Dopper cooling is possible when using a more advanced scheme~\cite{Gokhroo2011,Landini2011}. For the case of \39K, this is realized as follows: At the beginning of the molasses phase, the repumping light power is abruptly lowered to $5\%$ of its initial value and tuned on resonance. At the same time the trapping light is detuned to $-0.5~\Gamma$. Within the following \unit[9]{ms}, its power is linearly ramped down by $50\%$ and the detuning is simultaneously increased to $-2.3~\Gamma$. These molasses procedures are simultaneously performed for \39K and \87Rb and obtain temperatures of \unit[117]{\textmu K} and \unit[35]{\textmu K} respectively.

To obtain efficient transfer into a magnetic potential, it is necessary to optically pump both species to the $\ket{2,2}$~state. For this purpose, a homogeneous offset field of \unit[15]{G} is applied and $\sigma_+$ polarized light close to the $\ket{2}$ to $\ket{2'}$ transition is applied to both atomic species for a duration of $\unit[1]{ms}$. Additionally, repumper light is applied to prevent atom loss to the ground state $\ket{1}$ manifold. The optical pumping increases the number of transferred atoms by $50\%$.
	
After optical pumping, both species are transferred into a magnetic quadrupole trap. The current through the quadrupole coils is abruptly increased to \unit[15]{A} to catch the atoms and then ramped up in \unit[50]{ms} to \unit[45]{A} resulting in a gradient of \unit[196]{G/cm} (always given for the strongest, vertical direction). During this procedure the atoms change their spatial spin orientation while maintaining their spin state. The procedure was optimized empirically to obtain maximal transfer with minimal additional heating.

Finally the atoms are moved to the science chamber within \unit[1.2]{s} by mechanically moving the quadrupole coils. There they are loaded into a stationary magnetic quadrupole potential within \unit[800]{ms}. This is achieved by lowering the transport gradient while increasing the stationary coil gradient to \unit[350]{G/cm}. During this transfer, the atomic clouds move a transverse distance of \unit[4.5]{cm} into the L-shaped science cell. At this point, $2\EE6$ \39K and $6\EE8$ \87Rb atoms at a temperature of \unit[370]{\textmu K} are available.

\subsection{Magnetic trapping and evaporative cooling}
\label{magnetictrap}
	
\begin{figure}[tbp]
	\centering
    \includegraphics[width=8.6cm]{\folder 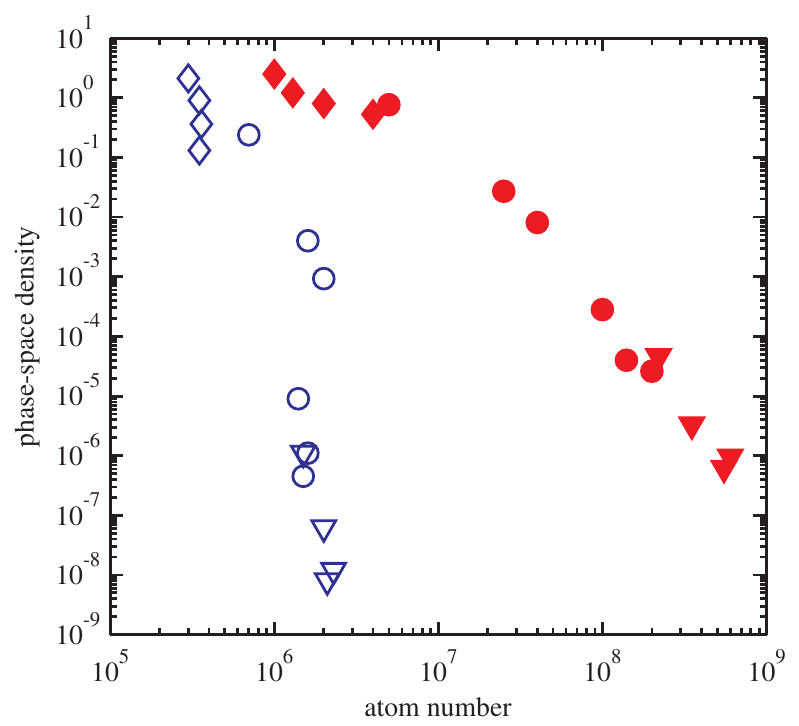}
		\includegraphics[width=6cm]{\folder 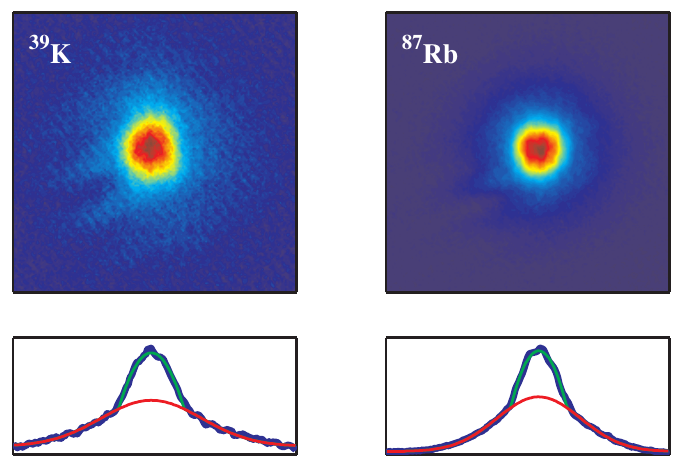}
	\caption{Phase-space density during the sympathetic cooling sequence as a function of the number of atoms. (top) While the number of \87Rb atoms (red solid symbols) is reduced to lower the temperature and raise the phase-space density, the number of \39K atoms (blue open symbols) stays approximately constant. The joint state preparation and dipole trap loading leads to a small loss in both species, visible as a kink at phase-space densities around \unit[0.5]{}. The symbols indicate measurements taken in the quadrupole trap (triangles), QUIC trap (circles) and dipole trap (diamonds). (middle) Typical images of BECs of \39K (left) and \87Rb (right) after \unit[15]{ms} and \unit[17]{ms} of ballistic expansion, respectively. The images have a size of \unit[450]{\textmu m} by \unit[450]{\textmu m}. (bottom) Column sums of the images (blue) and bimodal fits to the thermal (red) and BEC distribution (green).}
	\label{fig:PSDvsN}
\end{figure}

The initial sympathetic cooling of \39K with \87Rb is performed in the stationary quadrupole trap, which provides high thermalization rates~\cite{Ketterle1996}. Since \39K and \87Rb have the same linear Zeeman splitting, radio frequency transitions to untrapped $m_F$ states would remove both \39K and \87Rb atoms. Therefore, the rubidium atoms are evaporatively cooled by using a microwave field which transfers them from the trapped state $\ket{2,2}$ to the untrapped state $\ket{1,1}$. Due to the spatial overlap of the two clouds, this leads to sympathetic cooling of the \39K atoms as shown in Fig.~\ref{fig:PSDvsN} (triangles). Experimentally the microwave frequency is ramped from $\nu_{0}~+~135$~\unit{MHz} to $\nu_{0}~+~30$~\unit{MHz} within \unit[5]{s} while the quadrupole gradient is simultaneously reduced to \unit[220]{G/cm}. Here $\nu_{0}$ corresponds to the unperturbed transition frequency from $\ket{1,0}$ to $\ket{2,0}$. At this point the evaporation is stopped, since losses due to collisions with rubidium atoms in the $\ket{2,1}$~state become detrimental. This produces rethermalized samples at a temperature of \unit[180]{\textmu K}.  

Further cooling is realized in a harmonic magnetic potential~\footnote{In first experiments, a hybrid trap~\cite{KleineBuning2010,KleineBuning2011} was used in an attempt to sympathetically cool potassium with rubidium. However, the initial microwave evaporation of rubidium in the quadrupole trap has the drawback of producing hot atoms in the $F=2,m_F=1$~state. This ultimately prevented the loading of sufficient atoms in the dipole trap to produce BEC.}, which does not suffer from $m_F$-changing Majorana losses, and allows for the suppression of losses due to atoms in the $\ket{2,1}$~state. The atoms are transferred into a Quadrupole Ioffe-Pritchard configuration (QUIC) trap~\cite{Esslinger1998}, by increasing the current through an additional coil within \unit[1.4]{s}. This results in a magnetic potential with axial and radial trapping frequencies of $\nu_a=\unit[17.4]{Hz}\ (\unit[26]{Hz})$ and $\nu_r=\unit[195]{Hz}\ (\unit[291]{Hz})$ for \87Rb (\39K) at an offset field of \unit[1]{G}. After loading, we obtain \unit[$1.5\EE6$]{} \39K and \unit[$2\EE8$]{} \87Rb atoms at a temperature of \unit[83]{\textmu K} in the QUIC trap, corresponding to a phase-space density of \unit[$4.5\EE{-7}$]{} and \unit[$2.6\EE{-5}$]{} respectively.
	
The microwave evaporative cooling is continued in the QUIC trap for \unit[28]{s}. To remove atoms accidentally transferred into the $\ket{2,1}$~state, a second microwave is used, which is resonant with the transition from $\ket{2,1}$ to $\ket{1,1}$ at the center of the trap. The evaporation is stopped at a frequency of $\nu_{0}+\unit[2.54]{MHz}$. This produces samples at a temperature of \unit[800]{nK} containing $7\EE5$ \39K and $5\EE6$ \87Rb atoms, corresponding to a phase-space density of $0.24$ and $0.76$, respectively  as shown in Fig.~\ref{fig:PSDvsN}. If only \87Rb is used, further evaporation at this point allows the production of almost pure \87Rb BECs containing about $5\EE{5}$ atoms~\footnote{This also allowed for the  production of \41K BECs via sympathetic cooling in the magnetic trap.}.

Due to the negative background scattering length of \39K, further evaporation is performed in an optical dipole trap where the scattering length can be tuned using magnetic fields. The endpoint of the microwave evaporation was chosen to optimize the loading efficiency of \39K by matching its temperature to the depth of the dipole potential.

\subsection{Dipole trapping and state preparation}
\label{dipoletrap}
	
An optical dipole potential is required to exploit the tunability of both the intraspecies scattering length of \39K and the interspecies scattering length of \39K and \87Rb. In particular, the negative intraspecies background scattering length of \39K mandates tuning of the scattering length. To access the favorable combination of Feshbach resonances shown in Fig.~\ref{fig:Scattering}, it is necessary to prepare both species in the $\ket{1,-1}$~state.

The experiments are performed in a far red-detuned, crossed-beam dipole trap at a wavelength of \unit[1064]{nm}. This light is derived from a fiber amplifier(\textit{Nufern PSFA-1064-50mW-50W-0}) seeded by a narrow bandwidth monolithic ring laser (\textit{Coherent Mephisto}). The two required dipole laser beams are delivered to the experiment by high-power optical fibers and the output powers are stabilized by acousto-optic modulators in servo loops which also provide a \unit[220]{MHz} frequency difference between the beams. The two beams are aligned perpendicular to each other in the horizontal plane and are both focused to waists of \unit[100]{\textmu m} at the location of the atoms. This provides an attractive harmonic potential, which is largely independent of the spin state of the atoms. At a power of \unit[1.1]{W} per beam, vertical and radial trapping frequencies of $\nu_z=\unit[139]{Hz}\ (\unit[199]{Hz})$ and $\nu_r=\unit[102]{Hz}\ (\unit[142]{Hz})$ for \87Rb (\39K) are obtained.

\begin{figure}[htbp]
	\centering
		\includegraphics[width=8.6cm]{\folder 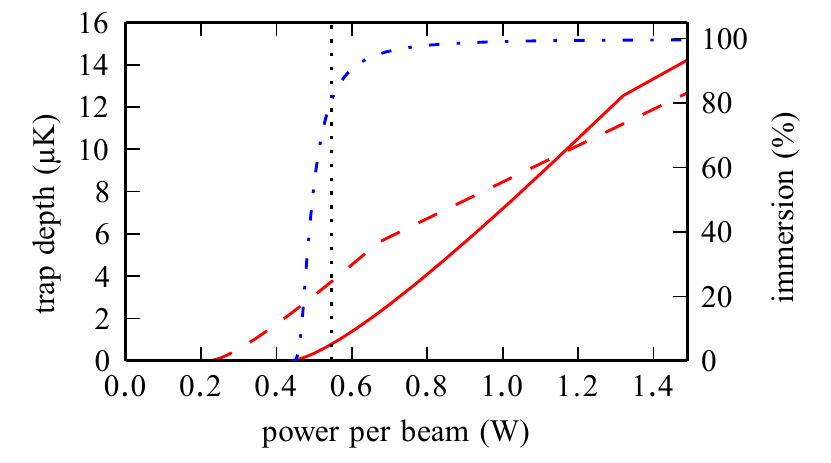}
	\caption{Trap depths as a function of the dipole trap power. The trap depths for \39K (red dashed line) and \87Rb (red solid line) show characteristic kinks where the trap becomes shallowest in the vertical direction due to gravity. The immersion of \39K (blue dash dotted line) in \87Rb is strongly reduced when the trap becomes weak for \87Rb and the differential gravitational sag becomes relevant. The dotted vertical line indicates the expected power to obtain BEC of \39K for $2\EE5$ atoms and $\eta=3$, where $\eta$ denotes the ratio between trap depth and temperature of the sample.}
	\label{fig:100mu-trap}
\end{figure}

Figure~\ref{fig:100mu-trap} shows the depth of the dipole trap depending on the beam power. The waist was chosen such that the immersion of \39K in \87Rb  is sufficient to ensure thermalization at the point where condensation of \39K is expected to set in (vertical line). This choice places the crossing point between the depths for \39K and \87Rb at \unit[10]{\textmu K}, allowing for efficient loading of atomic samples at temperatures up to \unit[1]{\textmu K}. Hence evaporation above \unit[1.1]{W} primarily leads to loss of the minority species \39K, contrary to the aim of sympathetic cooling. Below \unit[1.1]{W} sympathetic cooling of \39K with \87Rb can work efficiently. If the power is decreased below \unit[0.45]{W} per beam, \87Rb is not trapped anymore and pure \39K samples can be prepared.

Experimentally, atoms are loaded into the dipole trap as follows. First the QUIC trap is decompressed to trap frequencies of $\nu_a=\unit[13.5]{Hz}\ (\unit[20.2]{Hz})$ and $\nu_r=\unit[44.1]{Hz}\ (\unit[65.9]{Hz})$ for \87Rb (\39K) in \unit[1.2]{s}, resulting in a temperature of \unit[400]{nK}. Then, the current in the trap coils is linearly turned off over \unit[1]{s}, while the power in the two dipole beams is ramped up to \unit[0.9]{W}. Thus, \unit[$4\EE5$]{} \39K atoms and \unit[$4\EE6$]{} \87Rb atoms with a temperature of \unit[1]{\textmu K} are obtained in the dipole trap, corresponding to a phase-space density of \unit[0.1]{} and \unit[0.5]{} respectively (Fig.~\ref{fig:PSDvsN}). 

To address the Feshbach resonances shown in Fig.~\ref{fig:Scattering}, the atoms in both species have to be transferred from the $\ket{2,2}$ to the $\ket{1,-1}$~state. This is achieved in a three step process. First, a rapid adiabatic passage transfers both species simultaneously into the $\ket{2,-2}$~state by sweeping a radio frequency from \unit[6.0]{MHz} to \unit[8.2]{MHz} in \unit[2]{ms} at a homogeneous background magnetic field of \unit[10]{G}. This is possible, since the difference in the quadratic Zeeman splitting  between \39K and \87Rb is sufficiently small. To transfer the \87Rb atoms into the lower hyperfine state a single resonant microwave $\pi$-pulse with a duration of \unit[6]{\textmu s} at a field of \unit[1.8]{G} is used. Finally, the \39K atoms are transferred to the $\ket{1,-1}$~state by another rapid adiabatic passage, implemented by sweeping the background field from \unit[15]{G} to \unit[3.5]{G} while irradiating the sample with a radio frequency of \unit[450]{MHz}. The choice of technique for each of these steps is based on the available frequency tunability and power.

\begin{figure}[htbp]
	\centering
		\includegraphics[width=8.6cm]{\folder 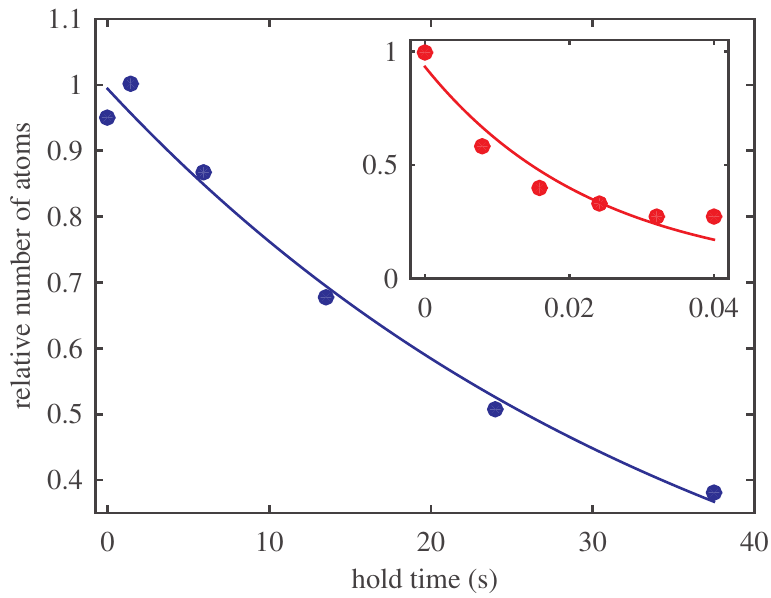}
	\caption{Decay of \39K in the presence of \87Rb in different state combinations in the dipole trap. When both \87Rb and \39K are transferred to the $\ket{1,-1}$~state \39K lifetimes of \unit[40]{s} are obtained. (inset) If \87Rb remains in the $\ket{2,-2}$~state, strong losses due to hyperfine changing collisions lead to \39K lifetimes of \unit[20]{ms}.}
	\label{fig:lifetimes}
\end{figure}

Note that it is vital to transfer rubidium to the $\ket{1,-1}$~state first, since severe losses are observed when the order is reversed. Figure~\ref{fig:lifetimes} shows the lifetimes for different state combinations in the dipole trap. When both species are in the target state, lifetimes on the order of \unit[40]{s} are observed. If however only \39K is in the $\ket{1,-1}$~state while \87Rb remains in the $\ket{2,-2}$~state a rapid decay leads to lifetimes on the order of \unit[20]{ms}. We attribute this loss to hyperfine changing collisions, in which a \39K and a \87Rb atom exchange spin states leading to the release of a kinetic energy corresponding to the difference in hyperfine splitting. If the state population is reversed, the process is energetically forbidden and hence long lifetimes are observed. A similar process has been observed in ultracold mixtures of \41K and \87Rb gases~\cite{Thalhammer2008}. 

In our experiments the duration of the state preparation process is comparable to the rapid decay time and thus a transfer in the incorrect order leads to severe losses. However, a transfer in the correct order yields atom numbers and lifetimes that are sufficient for the final evaporation steps to dual-species BEC.

\subsection{Feshbach tuning and BEC production}
\label{dualBEC}

The final evaporative cooling steps towards dual condensation require tuning of the scattering lengths by using the Feshbach resonances shown in Fig.~\ref{fig:Scattering}. Generally stable \39K condensates can be produced in the entire region between the two intraspecies Feshbach resonances at \unit[33.6]{G} and \unit[162.35]{G}. The interspecies Feshbach resonance at \unit[117.56]{G} divides this region in two parts where dual condensates of \39K and \87Rb can be created. Within our experiments, the production of dual-species BECs in both of these regions was feasible with minor adjustments of the final evaporation ramp. In practice, we typically create dual-species BECs at \unit[142.5]{G}, in the center of the region at higher field, since this provides a good starting point for further experiments (see section~\ref{miscibility}). 

We use the same coils which previously produced the stationary quadrupole potential but now in Helmholtz configuration to address this field. Since we require high field stability we have implemented a battery powered current supply based on a precision current transducer (LEM IT200-S Ultrastab). In addition, we actively stabilize the background magnetic field in the vicinity of our experimental chamber by controlling the current in two large compensation coils. These measures result in magnetic field fluctuations of \unit[5]{mG} determined by spectroscopic measurements on hyperfine transitions in \87Rb at \unit[117]{G}.

The final sympathetic cooling sequence is initiated by raising the magnetic field to \unit[142.5]{G}. Immediately afterwards, the evaporation is started by simultaneously decreasing the intensity in both dipole beams. Within \unit[16.5]{s}, the power per beam is decreased in 3 linear steps from \unit[0.9]{W} to \unit[480]{mW}. As expected, we first observe the condensation of \39K and shortly afterwards the condensation of \87Rb. Figure~\ref{fig:PSDvsN} shows typical images of these BECs.

The efficiency of the evaporation is characterized by $\gamma = - \frac{d (\ln D)}{d (\ln N)}$, where $D$ is the phase-space density and N the number of atoms. In the magnetic trap this yields an efficiency of $\gamma = 3.15$ for the evaporation of \87Rb while it sympathetically cools \39K. For \39K  $\gamma = 18$ is obtained, which demonstrates the efficiency of sympathetic cooling in the absence of strong losses. The evaporation efficiency of \87Rb is slightly lower in the dipole trap, however only a very small part of the evaporation is carried out in these conditions before reaching dual condensation.

At the end of each experimental run, absorption images of both species are taken after free expansion. After release from the dipole trap, the atomic clouds expand at the final magnetic field of that particular run for \unit[6]{ms}. Typically, the non-linear part of the expansion lies within this time and the atomic clouds expand ballistically afterward. The magnetic field is then turned off, and the field decays to a background value of \unit[1]{G} within another \unit[6]{ms}. To avoid effects due to this decay, the absorption image of \39K is taken after a total time of flight of \unit[15]{ms}. The image of \87Rb is taken another \unit[2]{ms} later due to the limited readout time of the CCD camera. Both images are taken with light resonant to the $\ket{2}$ to $\ket{3'}$
transition. Repumping light is applied \unit[200]{\textmu s} prior to the imaging light. If we want to characterize the state composition e.g. to evaluate the performance of state preparation, we turn on an inhomogeneous magnetic field in the first \unit[5]{ms} of the expansion. This fully separates the different $m_F$-components and allows for the detection of small admixtures in other states.

\section{Tunability of interactions}
\label{interactions}

\subsection{Feshbach resonance position}
\label{Feshbach}

Most experiments with heteronuclear mixtures rely on precise tuning of the interspecies interaction strength. Our experiments are performed with the Feshbach resonance at \unit[117.56]{G} shown in Fig.~\ref{fig:Scattering}. This resonance was previously characterized with good precision within a larger survey of Feshbach resonances~\cite{Simoni2008}. However, the precision of the resonance position was not sufficient for scattering length tuning in present work and the width of the resonance was only known from theoretical calculations. We therefore present a precise measurement of these quantities in the following.

\begin{figure}[htbp]
	\includegraphics[width=8.6cm]{\folder 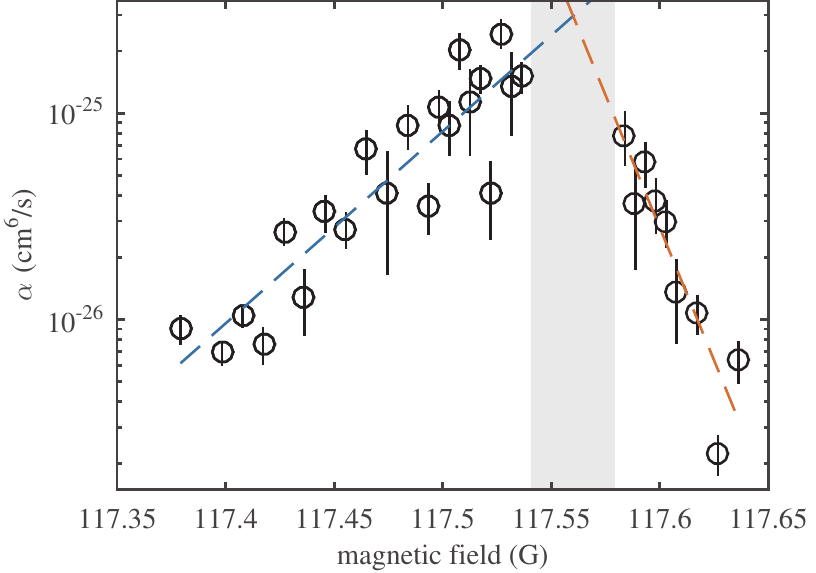}
	\caption{Three-body recombination rate near the Feshbach resonance at \unit[117.56]{G}. The dashed lines represent exponential fits to the data on either side. A region of fast decay at the center of the resonance was excluded.}
	\label{fig:resonance_position}
\end{figure}

The three body loss coefficient $\alpha$ was recorded for mixed samples to measure the resonance position. The final evaporation was carried out at \unit[119.6]{G} and the evaporation was stopped slightly before condensation. Then the clouds were adiabatically compressed by increasing the power to \unit[850]{mW} per beam and the field was set to its target value and held there for a variable duration before measuring the atom number in each species. A numerical fitting procedure was used to extract the three body loss coefficient. Figure~\ref{fig:resonance_position} shows these loss coefficients as a function of the magnetic field. To determine the position of the Feshbach resonance, we fit exponential functions to both sides and extract a position $B_0=\unit[117.56(2)]{G}$~\cite{Roy2013}. The uncertainty of \unit[20]{mG} corresponds to the excluded region of fast decay close to the Feshbach resonance.

\begin{figure}[htbp]
  \includegraphics[width=8.6cm]{\folder 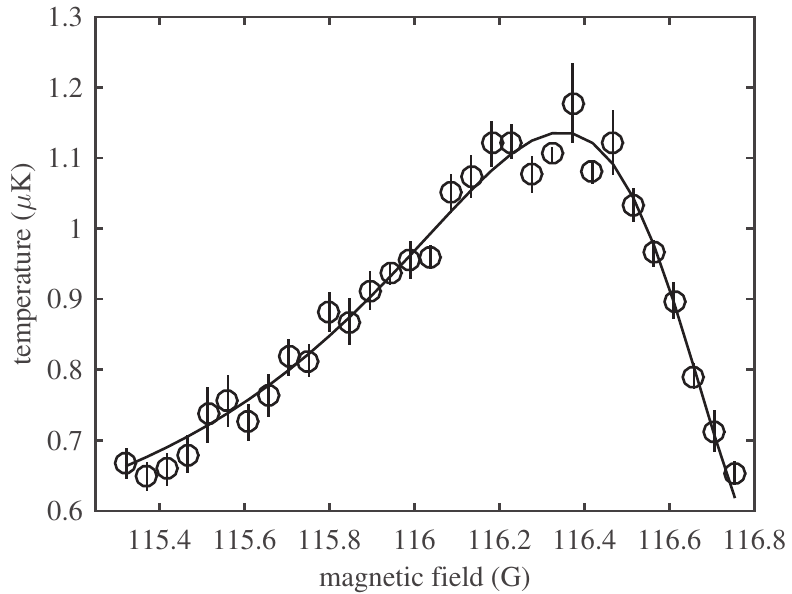}
	\caption{Temperature of \39K after sympathetic cooling with \87Rb close to the zero crossing of the interspecies scattering length. At the crossing sympathetic cooling is ineffective and \39K remains at a higher temperature. A fit to the data yields a field of \unit[116.35(3)]{G} for the position of the zero crossing~\cite{Thalhammer2008}.}
	\label{fig:resonance_width}
\end{figure}

The width of the resonance is determined by the distance between its center and the adjacent zero crossing of the scattering length. We have measured the position of this zero crossing by examining the efficiency of sympathetic cooling. Directly after the state preparation procedure, the magnetic field is set to a target value in the vicinity of the zero crossing and evaporation is started. This evaporation is stopped at a power of \unit[650]{mW} per beam and the temperature of \39K is measured (see Fig.~\ref{fig:resonance_width}). Since primarily \87Rb is evaporatively cooled and the sympathetic cooling fails when the interspecies scattering length vanishes, the temperature maximum of \39K indicates this point. We thus obtain a width of \unit[1.21(5)]{G} of the resonance, which is in agreement with previous theoretical results~\cite{Simoni2008}.

This data provides us with a precise model of the Fesh\-bach resonance, required to tune the interspecies scattering length in future experiment. Based on our magnetic field stability of \unit[5]{mG} we can thus access scattering length up to $\unit[\pm 1000]{}$ Bohr radii, if we allow for relative errors up to \unit[10]{\%}.

\subsection{Miscibility measurements}
\label{miscibility}

Dual-species mixtures with tunable interactions allow for the investigation of the phase transition from a miscible to an immiscible state. In the immiscible state the repulsive interaction between the two components is sufficiently large, such that the ground state of the system is phase separated. This transition is characterized by the miscibility parameter $\Delta=\frac{g_{11}g_{22}}{g_{12}^2}-1$~\cite{Edmonds2015} where the interaction strengths are $g_{ij}=2\pi\hbar^2 a_{ij} (m_{i}+m_{j})/(m_{i}m_{j})$. The indices refer to the two species with respective masses $m_{i}$ and scattering lengths $a_{ij}$. If this parameter is negative, $\Delta<0$, the mixture is immiscible while it is miscible in the positive case $\Delta>0$. The transition between these regimes was previously observed in Bose-Bose mixtures in different spin states~\cite{Tojo2010,Nicklas2011}, Bose-Bose mixtures of two isotopes of rubidium~\cite{Papp2008} and in heteronuclear Bose-Fermi mixtures~\cite{Ospelkaus2006,Zaccanti2006}. Recently, the availability of more Bose-Bose mixtures allowing for the creation of dual condensates has led to renewed theoretical interest. This has led to models~\cite{Edmonds2015} including all interactions between the condensed and thermal components of both species. 

\begin{figure}[htbp]
	\centering
	\includegraphics[width=8.6cm ]{\folder 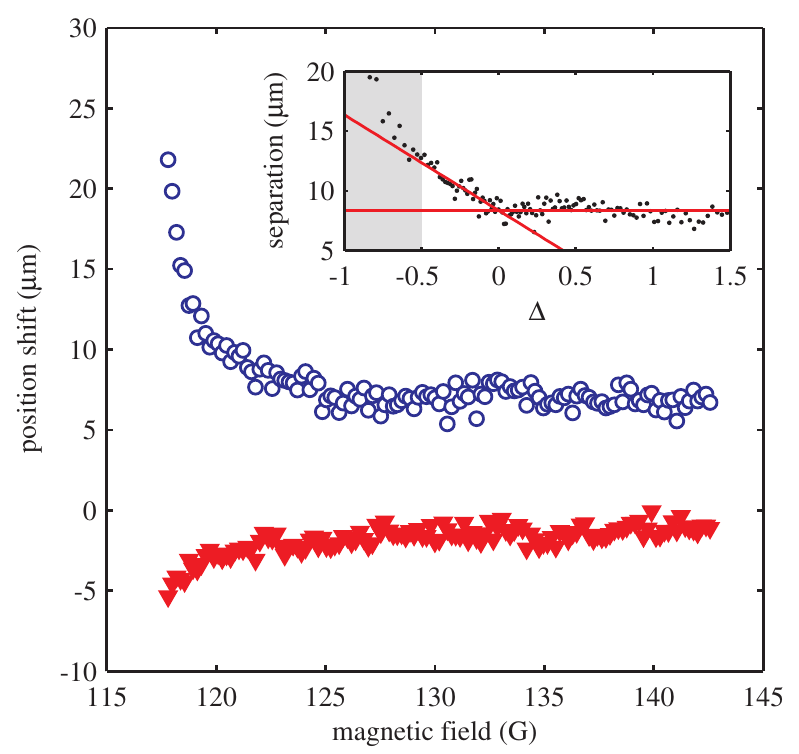}
	\includegraphics[width=6cm ]{\folder 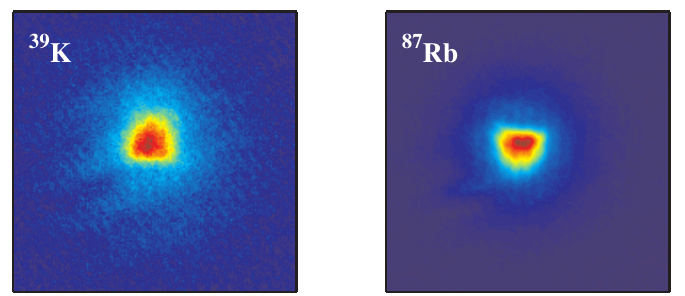}
	\caption{Shift in position of the two species after expansion. (top) Both clouds are systematically displaced from the single-species position. This displacement increases for large interaction strengths at magnetic fields close to the Fesh\-bach resonance. (inset) Separation of the two clouds as a function of the missibility parameter. An empirical linear fit to the data allows for the determination of the background scattering length (see text). (bottom) Sample images of \39K and \87Rb (size \unit[450]{\textmu m} by \unit[450]{\textmu m}) after a free evolution time of \unit[15]{ms} and \unit[17]{ms} respectively at a field of $\approx\unit[118]{G}$. }
	\label{fig:mixing}
\end{figure}

We have characterized the transition region from miscible to immiscible, to highlight the features of the \39K and \87Rb mixtures. A precise understanding of this transition in mass imbalances mixtures is important for numerous experiments utilizing the tunability of the interaction strength. In particular phase separation can modify effects 
expected due to few particle~\cite{Zinner2014} or polaronic particle~\cite{Massignan2014a} physics. 

Dual BECs were created as described in section~\ref{dualBEC}. Subsequently the dipole trap power was ramped to \unit[1]{W} in \unit[200]{ms} to compress the atomic clouds. Then the field was slowly ramped to a final field in the miscible or immiscible region within \unit[300]{ms}. After a further hold time of \unit[10]{ms}, the samples were imaged. In the miscible region, density distributions according to the self similar expansion of BECs from a harmonic trap were observed. However, in the strongly immiscible region, crescent shaped distributions with \39K displaced above and \87Rb below their initial position were observed as shown in Fig.~\ref{fig:mixing}. We have evaluated these positions by determining the shift in center-of-mass position of both species with respect to their position in single species experiments. This eliminates the shifts caused by the gravitational sag and by releasing the clouds from slightly different magnetic fields. 

Figure~\ref{fig:mixing} shows the shift in position for both species as a function of the applied magnetic field. Close to the resonance, the increased repulsion between the two species leads to significant shifts. However, these shifts cannot be attributed to a separation of the two components in the trap. The gravitational sag leads to a relatively large separation and the two trapped BECs hardly overlap in Thomas-Fermi approximation. Thus we attribute the observed effect primarily to the repulsion during the time-of-flight expansion. The full modeling of the process is an ongoing project.

The measured separation can be used as a novel method to determine the background scattering length $a_{bg}$. For this purpose the scattering lengths in the missibility parameter are substituted by their magnetic field dependent values.  We can thus plot the separation of the two clouds versus $\Delta$ for a trial background scattering length $a_{bg}$. We then fit this data with two linear dependencies for $-0.5<\Delta<0$ and $\Delta>0$ and obtain $a_{bg}$ by minimizing the error of these fits. The resulting fit is shown in Fig.~\ref{fig:mixing}~(inset) and yields a background scattering length of $a_{bg}=\unit[28.5]{a_0}$ which is comparable to the predictions for the adjacent $m_F$-states~\cite{Simoni2008}.

 \section{Summary}
\label{Summary}

We have presented the production of a new dual-species Bose-Einstein condensate and applied the tunablility of the interspecies scattering length in two experiments. The production process required a novel application of the dark-SPOT technique, which allowed us to considerably reduce the losses due to light-assisted collisions in the magneto-optical trapping phase and may be of benefit for other dual-species experiments. Moreover, it was shown that a particular state preparation sequence is necessary in the dipole trap to avoid losses due to hyperfine changing collisions in the dual-species condensation process. Preparation of both species in the $\ket{1,1}$~state enabled us to exploit a total of three Fesh\-bach resonances to obtain dual-species condensates and the ability to control the interspecies interactions. 

Precise tunability of the scattering length was used to perform precision spectroscopy of the interspecies Fesh\-bach resonance by evaluating the three-body loss coefficient. This allowed us to reduce the uncertainty on the position of the resonance by one order of magnitude. In addition, the width of the resonance was determined experimentally and importantly allows for precise tuning of the scattering length.

Finally, the transition from miscible to immiscible dual-species condensates with large mass imbalance was investigated, showing a clear effect during ballistic expansion. We show that this signal also allows for a model dependent determination of the interspecies background scattering length. These experiments provide a testing ground for theories~\cite{Edmonds2015} which include all interactions between the condensed and thermal components in a dual-species condensate. Moreover, a precise understanding of the behavior of mixed systems will be of interest for future precision experiments such as tests of the equivalence principle using \39K and \87Rb~\cite{Schlippert2014}.

\section{Acknowledgments}

We thank K. L. Lee and N. P. Proukakis for stimulating discussions on the effects of miscibility. We also thank the Lundbeck Foundation for financial support.

\bibliography{DualBEC}
\end{document}